\documentstyle[12pt,aasms4]{article}

\def\be{\begin{equation}}
\def\ee{\end{equation}}
\def\beq{\begin{eqnarray}}
\def\eeq{\end{eqnarray}}

\def\part{\partial}
\def\nn{\nonumber}

\def\la{\lambda}

\def\k{{\bf  k}}

\def\U{{\bf U}}
\def\B{{\bf B}}

\def\etal{{\it et al.\ }}
\def\b{{\bf b}}
\def\B{{\bf B}}

\def\v{{\bf v}}
\def\V{{\bf V}}
\def\OV{\overline{\V}}
\def\OB{\overline{\B}}

\begin{document}
\baselineskip=24pt
\begin{center}
{\Large\bf Numerical Analysis of Magnetic Field Amplification by Turbulence}
\bigbreak
{\large\bf by}
\medbreak
{\large\bf Hongsong Chou} \\
{\it Harvard-Smithsonian Center for Astrophysics, Cambridge, MA 02138,
U.S.A. \\ chou5@fas.harvard.edu}
\end{center}
\begin{abstract}
We apply a Fourier spectral numerical method to 3D incompressible MHD
turbulence with a magnetic Prandtl number $Pr \geq 1$. We examine the
processes by which an initially weak, large-scale seed magnetic
field and an initially weak, small-scale, impulse-like seed magnetic
field are amplified. We find that in both cases the magnetic energy
spectrum grows at all scales. The growth rates at different
amplification stages are analyzed. For a large-scale seed magnetic
field, the magnetic energy density grows as $\sim t^2$ for the first
few turbulence eddy turnover times, followed by a dynamic growth stage,
where nonlinear interactions between different scales of the turbulence
contribute to an exponential growth rate that is largely determined by
the turbulence eddy turnover time. For a seed magnetic field that is
initially set up at a small scale in the turbulence, during the
kinematic development stage, the growth rate of magnetic energy is
$\propto 1/\tau_{max}$, where $\tau_{max}$ is the eddy turnover time of
the smallest eddies of the turbulence. The kinematic growth stage is
followed by a dynamic growth stage, where nonlinearity plays important
role. During such dynamic growth stage, the growth rate of total
magnetic energy is determined by both the magnetic energy amplification
within the turbulence inertial range and that within the turbulence
dissipation range.
\end{abstract}
\keywords{MHD turbulence, Dynamo}

\section{Introduction}
Astrophysical magnetic field has often been observed on much larger
scales than the scales of astrophysical turbulence. The connection
between the generation of large-scale magnetic field and the turbulence
of much smaller scales has been contemplated by astrophysicists for many
decades. Mean-field electrodynamics (MFE, see Moffatt 1978 or Krause \&
R\"adler 1980), among other theories, employs a two-scale approach to
the problem. It suggests that the helical turbulent motions of
astrophysical plasma may align small-scale magnetic field so that an
observable, large-scale magnetic field can be formed. Because MFE is
largely a linear theory, its applicability to nonlinear phenomena such
as the solar or interstellar turbulence has been questioned ever since
the introduction of this theory. Some early criticism of MFE was
discussed by Piddington (1975). He argued that kinematic solar dynamo
theories do not account for the removal of the large amounts of flux
generated each solar cycle. Recent objections to dynamo action have
their root in the problem of small-scale magnetic fields. The
amplification of seed magnetic field in galaxies has been
considered by Kulsrud \& Anderson (1992). The magnetic Prandtl number,
defined as $Pr=\nu/\la$ where $\nu$ is the molecular viscosity and
$\la$ the magnetic resistivity, is much greater than 1 in the
interstellar medium of galaxies. Kulsrud \& Anderson predict
that the growth rate of an initially weak, small-scale magnetic field
in $Pr \gg 1$ systems will be $\propto 1/\tau_{max}$, where
$\tau_{max}$ being the eddy turnover time of the smallest turbulent
eddies. Because $\tau_{max}$ is very small in astrophysical turbulence,
the growth rate of the small-scale magnetic field will be large. They
then argue that because of such fast amplification of magnetic energy
at small scales, the turbulence may be reduced dramatically in a short
period of time, so that it is unable to amplify the magnetic field at
scales much larger than the turbulent dissipation scale. The recent
work by Kinney \etal (2000, see also Cowley 2000) also criticizes the
galactic dynamo theory by arguing that because the galactic plasma has
a very large magnetic Prandtl number, any small-scale seed magnetic
field will grow quickly to lock the velocity field in a shear motion
pattern, where the dissipation term in the momentum equation, $\nu
\nabla^2 \U$, is balanced by the Lorentz force term, $\B \cdot \nabla
\B$. Here $\U$ and $\B$ are the velocity field and the magnetic field.

This work is motivated by the studies of Kulsrud \& Anderson and Kinney
\etal We noticed that the work by Kulsrud \& Anderson can
be valid only for the kinematic development of the magnetic field, as back
reaction of the magnetic field on the velocity field was not considered in
their governing equations. Therefore, their prediction that the magnetic
field at large scales may not grow after the velocity field is quenched
at small scales needs examination. We also noticed that the
simulation by Kinney \etal is for 2D MHD, which can be very different
from the case of 3D MHD. Moreover, they focused their research on those scales
that are smaller than the velocity dissipation scales, and did not
include the inertial range of the turbulence. To further
understand the physics of magnetic field amplification by turbulence
and its application to 3D astrophysical systems, we carried out a
numerical study of incompressible 3D MHD systems. With our
numerical model, we study the amplification of initially weak,
large-scale and small-scale seed magnetic field by turbulence. Unlike
the theoretical analysis of Kulsrud \& Anderson, our numerical
study is not restricted to the kinematic development stage of the MHD
system. Rather, we can also study the effects of back reaction.
Unlike the numerical work by Kinney \etal (2000), we include
an inertial range in our numerical model, and study whether
or not the magnetic field within the inertial range will grow,
especially when the magnetic field beyond the turbulence dissipation
scale ($l_{v,D}$) grows and significantly modifies the velocity field
near $l_{v,D}$.

The structure of this paper is as follows: we introduce our numerical
model in section 2; sections 3 and 4 are devoted to the detailed
numerical analysis of magnetic energy spectrum development for a
large-scale seed field and a small-scale seed field, respectively; the
physical implications and applications of our numerical results are
discussed in section 5; conclusions and suggestions for future work are
given in section 6.  Our numerical study is different from those of
Brandenburg (2000) and Cho \& Vishniac (2000) in the following ways:
first, not like the work of Cho \& Vishniac, in most of our
simulations, we have a magnetic Prandtl number $Pr=3$, so that the
velocity dissipation scale is larger than the magnetic dissipation
scale; second, there is no scale separation as the one discussed in
Brandenburg (2000); third, although Brandenburg (2000) studied cases of
various magnetic Prandtl numbers, the initial conditions for the
simulation runs in that work are different from those in our work;
therefore, the physical processes considered in that work and ours are
different. 

In this work, we distinguish four scales. We denote the size of
turbulence energy containing eddies as $L$. $L$ is also called the
outer scale or the integral scale of the turbulence. We denote the dissipation scales of the
velocity field and the magnetic field as $l_{v,D}$ and $l_{b,D}$,
respectively. For turbulence of large kinetic Reynolds number, we
have $L \gg l_{v,D}$. In Fourier space, we introduce two wave numbers
that correspond to $l_{v,D}$ and $l_{b,D}$: $k_{v,D} \sim 1/l_{v,D}$
and $k_{b,D} \sim 1/l_{b,D}$. In some of our simulation runs, we have
$k_{b,D} > k_{v,D}$, i.e., the magnetic dissipation scale is smaller
than the velocity dissipation scale. This is consistent with a magnetic
Prandtl that is greater than 1. In other simulation runs, we have
$Pr=1$. Another scale is the so-called ``ensemble average scale'',
which is denoted by $\Gamma \gg L$ and over which we calculate averaged
quantities, $< \cdot >$ (or ${\overline {\mbox{  $\cdot$
}}}$). Finally, we denote the scale of the whole physical system as
$S$, which is the typical scale for the variations of averaged
quantities $< \cdot >$ (or ${\overline {\mbox{  $\cdot$
}}}$). What we mean by ``large scale'' in the following sections is the
scale $\sim S$. So we have the relation $S \gg \Gamma \gg L \gg
l_{v,D} \geq l_{b,D}$. In our model, there are two large-scale
quantities, $\OB$ and $\OV$. We assume both of these quantities are
constant, i.e., $S \rightarrow \infty$. We set our reference frame to
that moving at $\OV$ and henceforth omit terms of $\OV$. To aid our
discussion, we decompose the total magnetic field into two parts: $\B =
\OB + \b$, where $\b$ is the fluctuating component of $\B$. To make our
discussion easy to follow, throughout the paper we will use a few terms
and notations, which will be introduced as we proceed. They are also
listed in Table 1. Readers may refer to this table for further
clarification.  

\section{The Numerical Model}
We focus on the incompressible MHD equations, which are solved numerically
using the standard Fourier spectral method. The nonlinear terms are
evaluated by a pseudo-spectral procedure. Let {\bf B} and {\bf U} be
the magnetic field and the velocity field, respectively. Under an external
forcing term ${\cal F}$, the undimensionalized incompressible MHD
equations can be written as (with Einstein summation convention)
\be
\left( \partial_t - R_e^{-1} \nabla^2 \right) U_i = \partial_j
\left(-p\delta_{ij} - U_iU_j + B_iB_j \right) + {\cal F}_i,
\ee
\be
\left( \partial_t - R_m^{-1} \nabla^2 \right) B_i = \partial_j
\left(U_iB_j - B_iU_j \right),
\ee
\be
\partial_iU_i = \partial_jB_j = 0.
\ee
where $R_e$ and $R_m$ are the kinematic and magnetic Reynolds numbers
and defined as
\be
R_e=\frac{v_{rms}L}{\nu}, R_m=\frac{v_{rms}L}{\lambda}.
\ee
Here $v_{rms}$ is the root-mean-square of the velocity field, $L$ the
integral scale of the turbulence, $\nu$ the molecular viscosity,
$\lambda$ the magnetic resistivity. Note that we have already written
$B$ in units of $\sqrt{4\pi\rho}$ after we divide both sides of the
momentum equation and the induction equation by density $\rho$. If we
use a hat, $\wedge$, to denote discrete Fourier transform, and $\otimes$ to
denote convolution, the above equations in Fourier space are
\be
\left( \part_t + \nu k^2 \right) {\hat U}_j = P_{jl} \left
[ \mbox{i}k_m\left(-{\hat U}_l \otimes {\hat U}_m + {\hat B}_l \otimes
{\hat B}_m \right) + {\hat {\cal F}}_l \right],
\ee
\be
\left( \part_t + \la k^2 \right) {\hat B}_j = P_{jl} \left
[ \mbox{i}k_m\left({\hat U}_l \otimes {\hat B}_m -{\hat U}_m \otimes
{\hat B}_l \right) \right],
\ee
\be
k_m{\hat U}_m = k_j{\hat B}_j = 0.
\ee
Here $\bf P$ is the projection operator defined as $P_{jl} =
\delta_{jl} - \frac{k_j k_l}{k^2}$. In our simulation, we treat the
system as a cube $[0, 2\pi) \times [0, 2\pi) \times [0, 2\pi)$. The
Cartesian coordinate of a grid point can be written as $ x_l =
\frac{2\pi}{N}l, y_m=\frac{2\pi}{N}m, z_n=\frac{2\pi}{N}n,\mbox { for
$l,m,n=0,1,2,...,N-1$}$. A point in Fourier space has coordinates $k_s
= s, k_p=p, k_q=q,\mbox { for $s,p,q=-\frac{N}{2}, -\frac{N-1}{2}, ...,
\frac{N}{2} - 1$}$. 

Equations (5), (6) and (7) are numerically solved with the standard
Fourier pseudo-spectral method. Equations (5) and (6) are treated as ordinary
differential equations for ${\hat U}$ and ${\hat B}$. With the
projection operator {\bf P}, the divergence free condition (7) will be
satisfied for $t>0$ as long as ${\hat U}$ and ${\hat B}$ are divergence
free at $t=0$. All our simulations start from divergence free initial
conditions. We employ a second-order Runge-Kutta (RK2) method to advance
equations (5) and (6) in time. We can exploit the advantage of using
RK2 in the following two aspects. First, an integral factor can be
easily introduced with the transform
\be
{\cal U}_m (t) = {\hat U}_m (\k, t) e^{-\nu k^2 t}, {\cal B}_m (t) =
{\hat B}_m (\k, t) e^{-\la k^2 t}.
\ee
Second, aliasing errors can be reduced by introducing positive and
negative random phase shifts at the first and second stages of RK2,
respectively (Machiels \& Deville, 1998). The forcing term used
in our simulation is a combination of the one used by Chen \etal (1993a)
and the one used by Brandenburg (2000), with slight modification. Chen \etal
employed a forcing term that maintains the energy density values at
$k=1$ and $k=2$ so that the energy spectrum of velocity field always follows
Kolmogoroff $k^{-5/3}$ law at $k=1$ and $k=2$. In Brandenburg's
simulation, he injected helical waves of random phases at
$k=5$. Because our simulations start from an established hydrodynamic
turbulence, we first use the forcing by Chen \etal to obtain a fully
developed pure hydrodynamic turbulence. Such fully developed turbulent
velocity field is taken as the initial velocity field for the following
MHD turbulence simulation. For all the MHD turbulence simulation runs, the
forcing function has the form
\be
{\hat {\cal F}}(0.5 < |{\bf k}| \leq 1.5) = {\cal F}_c + {\cal F}_b.
\ee
That is, the force works only within the shell $S_1: 0.5 < |{\bf k}|
\leq 1.5$. Here ${\cal F}_c$ is a forcing term that is similar to the
one adopted by Chen \etal It is calculated by multiplying the velocity
components within shell $S_1$ by a factor, $\gamma > 1$,
so that before a new step of integration starts, the kinetic energy
density within this shell is reset to $E_1 = 0.24$. Phases of the
velocity components within the shell are not changed. This forcing is
equivalent as lengthening the velocity vector within shell $S_1$ by a
factor $\gamma -1$. Denote the increment of a velocity vector under
force ${\cal F}_c$ as $\delta{\bf v} = {\bf R} + \mbox{i}{\bf I}$, where
${\bf R}, {\bf I}$ are the real and imaginary parts of $\delta{\bf
U}$. In a few runs of our simulation, we need to inject kinetic
helicity into the turbulence. To do this with ${\cal F}_c$, we tune the
angle between $\bf R$ and $\bf I$ so that they remain perpendicular to
each other. Because kinetic helicity at $\k$ can be calculated as
$H(\k) = 2\k \cdot {\bf R} \times {\bf I}$, in doing so, we inject
kinetic helicity into the turbulence.

The forcing term ${\cal F}_c$ maintains the energy level at the forcing
scale so that the fluctuation in the energy development history can be
small; therefore, the growth stages of both the kinetic energy and the
magnetic energy can be studied carefully and accurately. However, this
force does not introduce random phases into the velocity field. To be
more realistic about the forcing in our simulation, we also use the
forcing term ${\cal F}_b$, derived from the forcing function
used by Brandenburg (2000), as a secondary forcing function to introduce
random phases into the velocity field. ${\cal F}_b$ has the form
\be
{\cal F}_b(\k) = {\cal F}_0 \frac{\k \times (\k \times {\bf {\hat e}}) - {\mbox
i}|\k| (\k \times {\bf {\hat e}})}{2k^2 \sqrt{1-(\k \cdot {\bf {\hat
e}}^2)/k^2}}  \cos (\phi (t)).
\ee
Here ${\cal F}_0<1$ is a factor adjusted at each time step so that the kinetic
energy density within shell $S_1$ fluctuates within 5\% of
$E_1$. ${\hat {\bf e}}$ is an arbitrary unit vector in Fourier
space. $\phi(t)$ is a random phase. Note that ${{\cal F}({\k})}^* =
{{\cal F}(-{\k})}$ so it is real, and it is helical in that ${\cal F}
\cdot \nabla \times {\cal F} = -k {\cal F}^2 < 0$, i.e., it has maximum
helicity. Because ${\cal F}_b$ is tuned in such way that it only
contributes to 5\% of the kinetic energy at the forcing scale, $F_b$
can be considered as a perturbation to ${\cal F}_c$. Therefore, the
advantage of using (9) as the forcing term is three fold: to avoid strong
fluctuations of kinetic and magnetic energy density with time, to
introduce random phases to the velocity field, and to maintain the kinetic
helicity at certain level.

In order to study the non-unit Prandtl number case, we adopt the
following hyper-viscosity and hyper-diffusivity: $-\nu_7 k^{14} {\hat
{\bf U}}$ and $-\la_7 k^{14} {\hat {\bf B}}$. The dissipation scale of
the velocity field is calculated with $k_{v, D} = \left(
\frac{\varepsilon_v}{\nu_h^3} \right)^{\frac{1}{6h-2}}$ (Machiels \&
Deville, 1998). Here $h=7$ is the order of the hyper-viscosity in our
simulations. $\varepsilon_v$ is the mean dissipation rate of the
velocity field and is calculated using the formula $\varepsilon_v =
2\nu_h \sum_k k^{2h} E_v(k)$, with $h=7$. $E_v(k)$ is the kinetic
energy spectrum. $k_{b, D}$ is defined in a very similar way as the one
$k_{v, D}$ is defined above. With $\nu_7 = 5.0 \times 10^{-16}$ and
$\la_7 = 3 \times 10^{-20}$ we have the dissipation scales of velocity
field and magnetic field as $k_{v,D} \approx 13$, $k_{b,D} \approx 28$,
respectively, so that the Prandtl number is $Pr = \left(
\frac{k_{b,D}}{k_{v,D}} \right)^{4/3} \approx 3$. In many other studies
(see Brandenburg 2000), the magnetic Prandtl number is defined as the
ratio of molecular viscosity to magnetic resistivity. Because we apply
hyper-viscosity in most of our simulations, the magnetic Prandtl number
used in this work can be considered as an ``effective'' magnetic
Prandtl number, i.e., a ratio inferred from the measured dissipative
cutoff wavenumbers. 

A 1D version of our code is used to solve the nonlinear Burger's
equation (see section 6.1 of Canuto \etal 1988) and the numerical
results match exactly the analytic results. If we impose a strong
uniform magnetic field and a small disturbance of velocity field, a
pair of Alf\'{v}en waves are numerically generated, both propagating along
the uniform magnetic field but in opposite directions. This is exactly
predicted by linearized incompressible MHD equations. Finite amplitude
MHD waves are also produced in our numerical simulations (see section
10.1 of Moffatt, 1978). Our code is also used to study hydrodynamic
turbulence with normal dissipation. With the forcing given by (9), the
flows reach a statistically stationary state in five to ten large eddy
turnover times. In Figure 1, we plot the kinetic
energy spectra of two different pure hydrodynamic turbulence simulation
runs with normal dissipation. The kinetic energy spectrum is calculated
as $E_{v}(k)=\frac{1}{2}\sum_{k-0.5}^{k+0.5} {\hat
v}(k^{\prime})^2$. The magnetic energy spectrum shown in next few
sections is calculated as $E_{b}(k)=\frac{1}{2}\sum_{k-0.5}^{k+0.5}
{\hat b}(k^{\prime})^2$. In Figure 1, $R_{\Lambda} = v_{rms} \Lambda /
\nu$ is the Taylor micro-scale Reynolds number. $\Lambda$ is the Taylor
micro-scale defined by $\Lambda = \sqrt{15 \nu v_{rms}^2/\epsilon}$, where
$\epsilon$ is the rate of dissipation of hydrodynamic kinetic energy
per unit mass (Chen \etal 1993b). For $R_{\Lambda} = 43$, the
exponential falloff starts around $k\sim6.5$, while for $R_{\Lambda}=70$,
the falloff starts around $k\sim11$. Both cases exhibit a Kolmogoroff
$k^{-5/3}$ inertial range. With a hyper-viscosity, the turbulence in
stationary state will have a energy spectrum that deviates from a
Kolmogoroff $k^{-5/3}$ inertial range. Rather, the spectrum will be
flatter than $k^{-5/3}$, as shown in numerous simulations
(Michiels \& Deville, 1998). We plot the kinetic energy spectrum with
the hyper-viscosity $\nu_7 = 5.0 \times 10^{-16}$ in Figure 1. The
spectrum in the inertial range follows a power law of $\sim
k^{-1.2}$. In Table 2, we have listed the parameters of all the
simulation runs that we have done for this work. The resolution of our
simulations is $(64)^3$.

\section{Amplification of a large-scale seed magnetic field by the turbulence}
The amplification of an initially weak, large-scale magnetic field is
studied by imposing a constant magnetic field along $y$-direction, $\OB =
{\overline B} {\hat {\bf y}}$ with ${\overline B}=0.0316$, into a
homogeneous, isotropic and stationary hydrodynamic turbulence under the
forcing of (9). The magnetic energy density associated with this
initial field is ${\overline E}_B=5\times10^{-4}$. Because the
turbulence will stretch $\OB$, magnetic field at $k\geq1$ will be
generated. If the magnetic field is weak, its back reaction on the velocity
field is small; therefore, the turbulence maintains its stationarity
until the magnetic field grows strong enough to alter the flows. The
growth rate of magnetic field due to the presence of a constant $\OB$
with negligible back reaction can be estimated as follows. The
induction equation can be written as 
\be
\part_t \b = (\OB \cdot \nabla) {\bf U} + \lambda \nabla^2 \b.
\ee
Therefore, shortly after the start of the simulation, i.e., during the
first few eddy turnover times, the magnetic energy spectrum within the inertial range can be
calculated as
\be
|{\hat {\bf b}}(\k)|^2 \approx t^2 {\overline B}^2 k^2 |{\hat {\bf
U}}(\k)|^2.
\ee
If the kinetic energy spectrum follows $|{\hat {\bf U}}(k)|^2
\propto k^{-p}$, we have $|{\hat {\bf b}}(\k)|^2 \propto
k^{-p+2}$. Figure 2 shows the energy spectra of the velocity field and
the magnetic field at $0.1 \tau_{eddy}$ after the start of
simulation. The kinetic energy spectrum changes little, while the
magnetic energy spectrum in the inertial range follows a power law of
$k^{0.77}$, as predicted. To show how fast
the magnetic energy grows in this stage, we calculate the following quantity
\be
\beta(t) = \frac{\Delta E_b}{\Delta t} \frac{1}{E_b} \sim \frac{2}{t}
\ee
for all $k$'s as well as the total magnetic energy. $\beta(t)$ can be
considered as an {\it instantaneous} growth rate. It is an explicit
function of and varies with time. From
equation (13), we find that $\beta(t)$ should not be a function of $k$,
and this is clearly
shown in the left panel of Figure 3, where we plot $\beta(t)$ at
different $k$'s. At different $k$'s, $\beta(t)$ is roughly the same,
and decrease with time. Near dissipation scales, $\beta(t)$ decreases
with time slightly faster than those within the 
inertial range, and this is due to the strong dissipation at large
$k$'s, i.e., the dissipation term in (11) becomes important. In the
right panel of Figure 3, we plot the averaged value (over $k$) of
$\beta(t)$, and it behaves like $\frac{2}{t}$ shortly after the start
of simulation, but decreases faster than $\frac{2}{t}$ as the strength
of $\b$ grows to be comparable with the strength of $\OB$. This
confirms equation (13). The deviation of the averaged $\beta(t)$ from
$\frac{2}{t}$ after $t=4$ becomes more and more prominent as the
nonlinear interaction between the velocity field and the growing
magnetic field becomes stronger and stronger. This is shown in Figures
4 and 5. From Figure 4, we find that the growth of the magnetic energy
density follows four stages. Stage 1 starts from the beginning of the
simulation till $t=3$, during which the magnetic energy density grows
as $t^{2}$. This is a linear stage in that the line stretching of $\OB$
by the velocity field contributes to most of the growth of the magnetic
energy, while the velocity field changes little. During stage 2 that
lasts from $t=4$ till $t=16$, it grows exponentially with a growth rate
$\beta = 0.1$. The velocity field is suppressed dramatically by the
end of stage 1 and at the beginning of stage 2. Figure 4 shows that the
kinetic energy density drops by 25\% during this period. After the
growth of magnetic energy enters the exponential stage, the loss of
kinetic energy slows down. Within stage 2, the velocity field loses 5\% of
its initial energy at $t=0$. The slowdown of the energy loss of velocity
field is due to the existence of a forcing, which injects energy at
$0.5 < |\k| \leq 1.5$ so that the $E_v( 0.5 < |\k| \leq 1.5)$ is
maintained at $0.240 \pm 0.012$. Because the nonlinear interaction
between the velocity field and the magnetic field is strongest near the
velocity dissipation scale $k_{v,D}$, the velocity at small scale is
suppressed most, which can be seen from Figure 5. The kinetic energy
density at $k_{v,D} = 13$ drops an order of magnitude from $t=0.31$ to
$t=9.42$. Therefore, the continuing growth of the total magnetic
energy during stage 2 should be attributed to the energy input from the
forcing scale and the velocity line stretching within the inertial
range. During stage 2, the dominant terms in the induction equation
within the inertial range is the line stretching term $\b \cdot \nabla
\U$, and this is because the fluctuating component $\b$ of the total
magnetic field has grown greater than the imposed $\OB$. The nonlinear
interaction between $\U$ and $\b$, and between $\b$ at different scales
provides a self-excitation of $\b$ that is independent of $\OB$ and is
capable of exponentially amplifying $\b$. Such an exponential growth
rate should be determined by the statistical properties of the velocity
field. In our simulation, the largest integral length that the largest
eddy circles around can be calculated as $L =
{2\pi\sum_k{k^{-1}E_v(k)}}/{\sum_kE_v(k)}$. The rms velocity of the
turbulence can be estimated as $v_{rms} = \sqrt{\frac{2}{3}
{\sum_kE_v(k)}}$. Within stage 2, we found that the temporal average of
$L$ (denoted by $<L>_t$) is $ <L>_t \approx 4.62$, and the temporal
average of $v_{rms}$ (denoted by $<v_{rms}>_t$) is $<v_{rms}>_t \approx
0.52$. Therefore, $1/\tau_{eddy} = <v_{rms}>_t/<L>_t = 0.11 \approx
{\beta}$, which is consistent with the theoretical prediction of Parker
(1979).

After the exponential growth stage, magnetic energy density growth
enters a near saturation stage from $t=16$ to $t=30$, where the growth
is further slowed down. During this stage, the magnetic energy at small
scales almost stops growing, while at large scales the magnetic field
continues to grow. For the saturation stage, magnetic energy density
fluctuates around 0.33, while the kinetic energy density fluctuates
around 0.4.  

The growth of magnetic structures can be seen from Figure 6, where we
plotted isosurfaces of the magnetic field strength,
$|\B|\equiv2.5\langle|\B|\rangle$. Here $\langle|\B|\rangle$ is the
spatial mean of the magnitude of the magnetic field. The isosurface at
a time shortly after the start of the simulation is shown in panel (a)
of Figure 6. At this time point, the magnetic field consists of mostly
small-scale structures. The structures shown in panel (a) occupy 3.5\%
of the whole system and are distributed quite evenly in space. However,
at a later stage of the development, the magnetic field develops
structures that span scales that are close to the forcing scale, $k
\approx 1$, in the system. This is shown in panel (b) at $t=31.4$ of
the saturation stage. Tube-like structures that span the whole
simulation box can be clearly seen.   

In our simulation of Run A, the kinetic helicity, calculated as $<\v
\cdot \nabla \times \v>$ as a function of time, has a value of $H_v = -0.13 \pm
0.04$. Most of the kinetic helicity comes from the forcing scale,  as
the force is helical. The normalized kinetic helicity is
${H_v}/\left({\omega_{rms} \cdot v_{rms}}\right) \approx {-0.13}/{0.74\cdot0.51} =
0.27 \pm 0.08$. Such helical flow will generate a dynamo
$\alpha-$effect and the $\alpha$ coefficient can be estimated as $\alpha \sim -\frac{\tau_{cor}}{3}<\v \cdot \nabla \times \v> \sim 0.05$, where
$\tau_{cor}$ is the correlation time of velocity field and is measured
to be $\tau_{cor} \sim 1.0$. A growth rate of the magnetic energy
density due to such a dynamo $\alpha-$effect can be estimated as
$\beta_{\alpha} \sim 2\alpha/L \sim 0.02$, which is smaller than the
growth rate in the second stage, $\beta = 0.1$. Therefore, it is not
clearly evident from this study that the dynamo $\alpha-$effect in the
moderately helical flow of our simulation is capable of driving the
exponential growth of magnetic field in stage 2. Cho and Vishniac
(2000) have done recent numerical simulations and also claim that a
dynamo $\alpha-$effect may play much less important roles in amplifying
magnetic field in turbulent flows than the turbulent line stretching
effect. One should also notice that the dynamo theory of mean-field
electrodynamics requires scale separation between the outer scale of
the turbulence and the scale of the physical system, while in our
simulation there is no scale separation; therefore, one should use
caution when applying our simulation results to the two-scale
discussion of the dynamo $\alpha-$effect. For two-scale approach to the
dynamo problem, the reader is referred to a recent work by Brandenburg
(2000), which presents a detailed account of large-scale magnetic field
amplification through dynamo $\alpha-$ and $\beta-$effects. 
 
\section{Amplification of a small-scale seed magnetic field by the turbulence}
In this section, we present the numerical analysis of a seed magnetic
field that is initially concentrated at a small scale, $k=20$. We start
from a fully developed turbulence and a magnetic impulse
\be 
E_b(k) = \left\{ \begin{array}{ll} e_0 & \mbox{for $k=20$,}
\nn\\ 0 & \mbox{otherwise.}  \end{array} \right.
\ee
Such an initial condition is consistent with the theoretical analysis by
Kulsrud \& Anderson (1992) on the amplification of weak small-scale magnetic
energy, which they called {\it magnetic noise}. The growth of magnetic
spectrum with $e_0 = 0.001$ (Run B) is shown in Figure
7. The initial impulse-like small-scale seed magnetic field is located
at $k=20$. After the simulation starts, the narrow impulse quickly
becomes broader and broader, extending to both larger and smaller
scales than $k=20$. It extends to large wave numbers (small scales) and
soon hits the magnetic dissipation scale $k_{b,D}=28$ and the energy is
removed by dissipation near $k_{b,D}$. The broadened, impulse-like,
seed field also extends to small wave numbers, transporting magnetic
energy to larger and larger scales. From $t=2.2$ to $t=19.0$, the portion
of the growing magnetic spectrum that extends from $k=1$ to $k_{v,D}$
follows a $k^{3/2}$ power law, indicated by a group of dashed lines in
Figure 7. Such $k^{3/2}$ profiles are reminiscent of the kinematic
analysis of the magnetic spectrum by Kulsrud \& Anderson (1992). Note
that by Kulsrud \& Anderson, such $k^{3/2}$ profiles should extend into
the range beyond $k_{v,D}$ but before $k_{b,D}$ is reached. However,
because of the low resolution of our simulation, we do not have a long
range between $k_{v,D}$ and $k_{b,D}$; therefore, the magnetic energy
spectrum beyond $k_{v,D}$ can be strongly affected by the dissipation
process near the magnetic dissipation scale and does not exhibit the $k^{3/2}$
scaling. Future simulations with higher resolution should be able to
resolve such scaling within that range.  

To study the broadening of the initial magnetic impulse, we define the
following two terms
\be
C_b(k) = \sum_{k-0.5 \leq |{\bf p}| < k+0.5} {\hat {\B}}^{*}({\bf p}) \cdot
{\hat {\bf Q}}({\bf p}),
\ee
\be
L_b(k) = \sum_{k-0.5 \leq |{\bf p}| < k+0.5} {\hat {\B}}^{*}({\bf p}) \cdot
{\hat {\bf R}}({\bf p}).
\ee
Here
\be
{\bf Q} = -\U \cdot \nabla \B
\ee
is the convection term, and $-\B \cdot {\bf Q}$ is the energy per unit
time that is transferred away from magnetic field at one location to
other locations. Also,
\be
{\bf R} = \B \cdot \nabla \U
\ee
is the term representing the line stretching effect, and $\B \cdot {\bf
R}$ is the work done to the magnetic field by the velocity field. In Figure 8,
we plot $kC_b(k)$ and $kL_b(k)$ at different time points. At $t=0.2$,
panel (a) of Figure 8 shows that the magnetic energy of the impulse at
$k=20$ is being  transported to scales larger and smaller than $k=20$. In
panel (b) of Figure 8, the energy is being removed from $k=20$ to
larger and smaller scales. It shows that the line stretching effect is
generating magnetic structures at other scales. Panels (a) and (b)
together explain the broadening of the impulse shown in Figure
7. Notice that at $t=0.2$, the convection effect, i.e., $C_b(k)$, is
comparable to the line stretching effect, i.e., $L_b(k)$. However, from
panels of the right column of Figure 8, we find that the line
stretching effect soon becomes dominant over the convection effect,
which is shown at different times in the left column of Figure
8. $kL_b(k)$ is non-negative at all $k$'s, therefore always converting
kinetic energy to magnetic energy. The convection term at
different $k$'s has different signs at different times, meaning that
the magnetic energy can be transported into or out of certain scale of
the MHD turbulence. Panels (b), (d), (f) and (h) also show that the
line stretching effect is most prominent at small scales, that is, the
scales between the velocity dissipation scale, $k_{v,D}=13$, and the
magnetic dissipation scale, $k_{b,D}=28$. 

According to Kulsrud \& Anderson(1992), before the back reaction of
magnetic field is strong enough to alter the velocity field, the
exponential growth of magnetic energy from a magnetic impulse should be
attributed to the line stretching at the velocity dissipation scale, i.e.,
\be
\frac{d{\cal E}}{dt} = 2 \gamma {\cal E}
\ee
where $\gamma = 1/\tau_{max}$ is the inverse of the eddy turnover time
of velocity field at dissipation scale $k_{v,D}$. Recent work by
Chandran(1997) and Schekochihin \& Kulsrud(2000) modified $\gamma$ from
$1/\tau_{max}$ to $\epsilon/\tau_{max}$ and  $\epsilon \approx 0.6$ is due
to the non-zero correlation time of the velocity field. We have found such
kinematic exponential growth stage in our simulations, such as the
$e^{\beta_1 t}$ portion of the magnetic energy growth in Figure 9. To test if the growth
rate, $\beta_1$, in such exponential growth stage can be estimated according
to the theory of Kulsrud \& Anderson, we have measured this quantity from
our simulation results. In Figure 10, we plot the kinematic exponential
growth stage of the magnetic field, which extends from $t=2$ to $t=9$ for
Run C with an $e_0(k=20) = 10^{-5}$. The best fitting to $E_b(t)$ gives a
growth rate of $\gamma_b = 0.36 \pm 0.05$. We also calculated the growth
rate due to line stretching by the largest eddies,
$\gamma_{eddy} = 0.18 \pm 0.04$. The smallest eddy turnover time calculated
from our numerical results within this time range is
$\gamma_n = 0.69 \pm 0.12$. Therefore, in our simulation,
the modification factor due to non-zero correlation time is $\epsilon =
\gamma_b/\gamma_n \approx 0.52$, and such result is consistent with the
work by Chandran (1997) and that of Schekochihin \& Kulsrud (2000).

For a Kolmogoroff turbulence, the eddy turnover time at scale $k_{v,D}$
can be calculated as
\be
\tau_{Kol,D}=\tau_{eddy} \left( \frac{k_{v,D}}{k_{eddy}}
\right)^{-2/3}.
\ee
With $\tau_{eddy}$ and $k_{eddy} \sim 1/L$ from our simulation, we have
$\gamma_{Kol,D} = 1/\tau_{Kol,D} = 0.57 < \gamma_n = 0.69$. This is due to the
fact that the hyper-viscosity flattens the kinetic energy spectrum, hence
produces larger values of velocity near the dissipation scale than that in 
Kolmogoroff turbulence. However, the nature of the growing of the magnetic
field due to line stretching near the dissipation scale does not change
with the slightly flattened kinetic energy spectrum. This can be seen
from the work by Kulsrud \& Anderson (1992). The growth rate of magnetic
energy depends on the integral
\be
\gamma \propto \int k^2 J(k,0) dk
\ee
where $J(k,0)$ kinetic energy density at zero frequency. For the
velocity field in our simulation, we have $J(k,0) \sim k^{-1.2}$, thus
most of the contribution to $\gamma$ still comes from the smallest
eddies.

\section{Discussion}
Our simulations show that the interaction between fully developed,
constantly forced turbulence and an initially weak seed magnetic field
will always lead to the growth of magnetic field at different
scales of the turbulence. Given an initially weak, large-scale external
magnetic field, the emergence of magnetic energy at small scales is due
to the line stretching of $\OB$ by the velocity field. This is a
kinematic process, as the velocity field is not affected much by the
growing magnetic field. Within the inertial range, where dissipation is
negligible, the magnetic field is amplified mainly by the line
stretching term, i.e., the second term of equation (11). This kinematic
stage finishes when the nonlinear interaction between the growing
 magnetic field becomes comparable to the line stretching
term, and a dynamic growth stage follows. In other words, when the
terms $\U \cdot \nabla \b$ and $\b \cdot \nabla \U$ are of the same
order as $\OB \cdot \nabla \U$, (11) is not valid anymore, and the
self-excitation of the growing magnetic field within the inertial range
will dominate other amplification processes. This is shown clearly as
stage 2 in Figure 4, where the total magnetic energy grows
exponentially. And from Figure 5, we find that such an exponential
self-excitation stage happens after the magnetic energy densities at
different scales grow to be comparable to the magnetic energy density
of $\OB$. 

Several authors have been arguing that there might be a relation
between the strength of a large-scale magnetic field, $\OB$, and the
strength of the fluctuating component of the magnetic field,
$\b$. Krause and R\"adler (1980, chapter 7) derived a relation for 3D
MHD between ${\overline B}^2$ and $\langle b^2 \rangle$ in the form
\be
\langle b^2 \rangle = {\overline B}^2 \frac{\eta_T}{\eta},
\ee
where $\eta_T$ and $\eta$ are the turbulent diffusion coefficient and the
magnetic diffusivity, respectively. The governing equation they used in
their derivation of (22) was equation (11), and the nonlinear terms
${\bf Q}$, ${\bf R}$ in (17) and (18) were ignored. Our
simulations show that this is not a valid procedure, as terms like
${\bf Q}$ and ${\bf R}$ are of importance to the exponential
growth of the magnetic energy. In fact, from our simulations we
find that no matter how large or small the initial large-scale magnetic
field is, after $5-10$ eddy turnover times of the turbulence, a
statistically stationary magnetic energy spectrum will be formed. 

The amplification of the small-scale seed field, simulated in Runs B and
C, further shows that relation (22) is not necessarily true, as $\OB=0$
in these two runs. The growth of magnetic field at all scales (see
Figure 7) can be due only to the nonlinear interactions between the
velocity field and the magnetic field at different scales.

In the two simulation runs (B \& C) with a moderate magnetic Prandtl
number ($Pr = 3$), the magnetic energy within the inertial range,
$E_{\cal I}$, which is initially set to zero, grows to a steady state
in which $E_{\cal I} \geq E_{\cal D}$, where $E_{\cal D}$ is the
magnetic energy stored between the two dissipation scales, $k_{v,D}$
and $k_{b,D}$ (see Table 1 for more details). Some authors (Cowley
2000) argue that for large magnetic Prandtl number, the magnetic field at
small scales, i.e., $k_{v,D} \leq k \leq k_{b,D}$, grows so fast and
strong that it swamps the velocity field at all scales and the magnetic
field within the inertial range cannot grow at all. Our simulation,
which includes both the inertial range and the dissipation range, does
not support such picture. Instead, our simulation shows that $E_{\cal I}$ does
grow. The initial magnetic impulse at small scales spreads to all
scales between $k_{v,D}$ and $k_{b,D}$. The growth rate of magnetic
energy at these small scales is approximately the inverse of the eddy
turnover time of the smallest turbulence eddies, as predicted by
Kulsrud \& Anderson (1992). During this kinematic growth period, the
magnetic energy spectrum is peaked near $k_{b,D}$, and extends into the
inertial range with a profile of $\sim k^{3/2}$, providing seed
magnetic field in the inertial range. Such seed magnetic field does
grow, and the growth rate in the kinematic growth period is $\propto
1/\tau_{max}$. Let range $\cal D$ include all the scales between
$k_{v,D}$ and $k_{b,D}$ (see Table 1 for definitions of other
terms). As the magnetic field within range ${\cal D}$ continues to
grow, it starts to exert strong back reaction on the velocity field
near scale $k_{v,D}$ and suppresses it. This starts the dynamic growth
period of the magnetic field. The eddy turnover time $\tau_{max}$ of
the gradually suppressed velocity field increases, which reduces the
growth rate, $\beta_{\cal D} \sim 1/\tau_{max}$, of the magnetic field
within range ${\cal D}$. Because $\beta_{\cal D}$ is reduced, the back
reaction of the magnetic field near the scales $\sim k_{v,D}$ on the
velocity field is reduced, too. For the forced turbulence, line
stretching in range ${\cal I}$ (the inertial range, see Table 1) is
still in effect as the velocity field in this range is not completely
suppressed by the growing magnetic field. Therefore, the magnetic field
within the inertial range continues to grow at a rate smaller than
$1/\tau_{max}$ but not smaller than $1/\tau_{eddy}$. The growth rate of
the total magnetic energy during the dynamic growth stage is a
combination of the growth rates of $E_{\cal I}$ and $E_{\cal D}$, as
shown as the $e^{\beta_2 t}$ stage of Figure 9. In Figure 11, we
compare the growth history of $E_{\cal I}$ and $E_{\cal D}$. Let ${\cal
E}=E_{\cal I}+E_{\cal D}$, $R_i = E_{\cal I}/{\cal E}$ and $R_d =
E_{\cal D}/{\cal E}$. Figure 11 shows that $E_{\cal I}$ starts from
$0$, and continues to grow until it dominates $E_{\cal D}$ after $t
\sim 40$. In previous numerical simulations (Kinney \etal 2000),
$E_{\cal I}$ was not considered at all, and the growth of $\cal E$ is
attributed only to $E_{\cal D}$. Figure 11 shows that $E_{\cal I}$ is
as important as $E_{\cal D}$ in the amplification process of the
magnetic field in turbulence, hence it must be included in numerical
studies.  

We also studied a relevant process in Run D. The initial seed magnetic
field is composed of two components: a large-scale magnetic field,
$\OB$, and a small-scale magnetic field $\B_{k=20}$ that is initially
concentrated at $k=20$. The growth of the magnetic energy spectrum given
such combined initial conditions is shown in Figure 12. For this
simulation run, we set the initial magnetic energy density of $\OB$
equal to that of $\B_{k=20}$, so that we can compare the contributions
of these two initial seed fields to the growth of the magnetic energy at
each $k$. Figure 12 shows that both $\OB$ and $\B_{k=20}$ contribute to
the amplification of magnetic field at each scale. For a certain $k$ within the
inertial range, these two contributions to the magnetic energy density
at this scale race against each other: the contribution from $\OB$
initially grows as $\sim t^2$, followed by an exponential
growth $\sim e^{t/\tau_{eddy}}$; the contribution from $\B_{k=20}$
grows initially as $\sim e^{t/\tau_{max}}$, followed by another
near-exponential growth with a smaller growth rate. The growth rate of
magnetic energy density within the inertial range is greater than both
of the growth rates in Run A and Run B, and this is because in Run D,
both the large-scale and small-scale magnetic fields provide seed field
within the inertial range. We believe this model can be applied to many
real astrophysical systems, for example, in the regions where supernova
remnants mix with ambient interstellar medium. Supernova remnants
usually carry small-scale magnetic field, while the ambient
interstellar medium can be threaded by large-scale magnetic field. Both
the initial large-scale magnetic field in the interstellar medium and
the small-scale seed magnetic field in supernova remnants will
contribute to the growth of magnetic field at different scales. It is
also possible that in the early evolution stages of galaxies, the seed
magnetic field may have components of both large and small scales. As we
have shown in above discussions, because the contributions of these two
components to the amplification of magnetic field at different scales
are different, we have to treat both of them in discussions of dynamo
action in galaxies.

Scale separation, which is discussed extensively in Brandenburg (2000),
is not considered in this work. What we mean by ``large scale'' in this
work is the scale $k \rightarrow 0$, while the ``large scale'' in
Brandenburg (2000) is $k \sim 1$. That is, there are no scales between
the large scale $S$ and the forcing scale $k \sim 1$ in this work. We
force the turbulence near $k=1$, which is near the size of the
simulation box, while the forcing scale of Brandenburg (2000) is
$k=5$. However, we did find the scale-by-scale growth of magnetic
energy in our simulation. In the first two cases of our
simulation (Runs A, B and C), in the kinematic development stage,
magnetic energy near the kinetic dissipation scale is always stronger
than the magnetic energy at different scales of the inertial
range. When the magnetic field is amplified, the effect of back
reaction first comes in near the kinetic dissipation scale and the
growth of magnetic energy near that scale first slows down. After the
slowing down of magnetic energy growth first appears near $k_{v,D}$,
the energy growth at the scale that is slightly larger than the
dissipation scale starts to slow down, followed by the energy growth
slowing down at even larger scales (or smaller $k$'s). Such slowing
down, which starts from $k_{v,D}$, continues from the dissipation scale
and moves from right to left in $k-$space, until the magnetic energy
near the forcing scale stops growing. Then, the development enters
fully saturated stage, with the magnetic energy at different scales
fluctuating around their saturated values. Such successive slowing down
of magnetic energy growth from the kinetic dissipation scale to the
forcing scale is due to the effect of back reaction of the growing
magnetic field on the velocity field, and such effect of back reaction
comes in scale-by-scale, starting from $k_{v,D}$. Note that this is
different from the energy growth near scales $k < 5$ in the work by
Brandenburg (2000, Fig. 18) in the following sense: because the author
is studying the dynamo $\alpha-$ and $\beta-$effects, scale
separation is necessary as the author compares the numerical results
with the prediction of theoretical two-scale approach to the
problem. In that work, the growth of magnetic field at $k<5$ is mainly
due to an $\alpha-$effect, through which the helical velocity field
aligns the magnetic field within the scales from $k \sim 5$ (i.e., the
forcing scale) down to $k_{v,D}$ (i.e., the kinetic dissipation scale)
in such way that the magnetic field at scale $k \sim 1$ is
formed. In this work, on the other hand, the growth and the saturation
of magnetic energy density at different scales is due to the interplay
between the magnetic field line stretching by the turbulent velocity
field and the magnetic back reaction on the velocity field. 

\section{Conclusions and future work}
We have carried out numerical simulations of 3D incompressible MHD with a
magnetic Prandtl $Pr \approx 3$. Both the initial large-scale seed
magnetic field and the initial small-scale seed magnetic field can be
amplified. For a large-scale seed magnetic field, the magnetic energy
density grows as $\sim t^2$ for the first few turbulence eddy turnover
times, followed by an exponential growth, of which the growth rate is
$\propto 1/\tau_{eddy}$. For a seed magnetic field at an initial input
small scale, during the kinematic development stage, magnetic energy
can be transported to all scales larger and smaller than the initial input
scale: near the magnetic dissipation scales magnetic energy is removed
by magnetic resistivity, while from the outer scale of the turbulence
to the initial input scale, the magnetic energy spectrum follows a
profile of $\sim k^{3/2}$. The measurement of the growth rate during
this kinematic process confirms the theoretical prediction by Kulsrud
\& Anderson (1992) that the kinematic growth rate is $\propto
1/\tau_{max}$ where $\tau_{max}$ is the eddy turnover time of the
smallest eddies of the turbulence. Entering the dynamic growth stage, the
growth of magnetic field at small scales exerts a strong back reaction on
the velocity field near the velocity dissipation scale. The suppression of the
velocity field slows down the growth of magnetic field between the
velocity dissipation scale and the magnetic dissipation scale. However, the
magnetic field within the inertial range of the forced
turbulence continues to grow. The magnetic field within the inertial
range grows to a steady state that has a profile $\sim
k^{-1}$ between the forcing scale and the magnetic dissipation
scale. The contribution to the total magnetic energy from the magnetic
field within the inertial range dominates that from the magnetic field
between the velocity dissipation scale and the magnetic dissipation
scales. For real astrophysical systems, the initial seed magnetic field may
have both a large-scale component and a small-scale component, and they
would both contribute to the growth of magnetic field at all turbulence scales.

Our simulations may suffer from the relatively low resolution ($64^3$);
therefore, the results may not completely applicable to very large
magnetic Prandtl cases. The low resolution forces us to adopt the
hyper-viscosity $\nu_7$ and the hyper-resistivity $\la_7$ in a few runs of our
simulation. Although the introduction of hyper-viscosity and
hyper-resistivity does not change the fundamental physics of our
discussion given above, the simulation results, such as our estimates
of the growth rates, can be slightly different from the results
obtained from simulation runs with normal viscosity and normal
resistivity, as we have discussed in Section 4. Nevertheless, our
simulations show at least that the inertial range of the turbulence
must be included in the discussion of how the growing small-scale
magnetic field affects the growth of magnetic field at all scales of
the turbulence. This is of paramount importance to the correct
understanding of astrophysical dynamo processes. In future work, we
expect to study the growth of magnetic field at all turbulence scales
with larger numerical resolutions. 

\acknowledgments
I am very grateful to Prof. George B. Field and Prof. Eric G. Blackman
for many stimulating discussions. I also thank an anonymous referee for
many valuable and insightful comments. I benefited from my
discussions with Axel Brandenburg, Ben Chandran, Steve Cowley, Hantao
Ji, Russell Kulsrud, Keith Moffatt and Jason Moran. I also want to
thank the organizers and all  participants of the Astrophysical
Turbulence Conference held at ITP of UC Santa Barbara, from May 8 to
May 12, 2000. 

\appendix

\clearpage

\begin{deluxetable}{ll}
\footnotesize
\tablecaption{Terms Used in This Paper and Their Physical Meanings}
\tablewidth{0pt}
\tablehead{ \colhead{What we call} & \colhead{What we mean} }
\startdata
large scale   & scale $\sim S$, which $\rightarrow \infty$ (or $k_S
\rightarrow 0$) in this work\nl
small scale     & a scale that is between $k_{v,D}$ and $k_{b,D}$ \nl
outer scale of the turbulence  & scale $\sim L$ \nl
forcing scale        & scale $\sim L$ \nl
inertial range or range ${\cal I}$    & scales between $L$ and $l_{v,D}$ \nl
range ${\cal D}$       & scales between $l_{v,D}$ and $l_{b,D}$ ($k_{v,D} < k <
k_{b,D}$)\nl
$E_{{\cal I}}$ & magnetic energy density of range  ${\cal I}$ \nl
$E_{{\cal D}}$ & magnetic energy density of range  ${\cal D}$ \nl
$E_{v(b)}(t)$ & history of kinetic (magnetic) energy density evolution \nl
$E_{v(b)}(k)$ & kinetic (magnetic) energy spectrum \nl
${\overline E}_B$ & magnetic energy density of $\OB$ \nl
$e_0(k)$ & magnetic energy density that is concentrated at scale $k$ \nl
\enddata
\end{deluxetable}

\clearpage

\begin{deluxetable}{cccllccc}
\footnotesize
\tablecaption{Measurements of various physical quantities in different
simulation runs}
\tablewidth{0pt}
\tablehead{ \colhead{Run} & \colhead{${\overline E}_B$} & \colhead{${e_0}(k)$}
& \colhead{$\nu$} & \colhead{$\la$} & \colhead{$Pr$} &
\colhead{$k_{v,D}$} & \colhead{$k_{b,D}$} }
\startdata
A   &  $5\times10^{-4}$ & $-$ & $\mbox{hyper, }\nu_7=5\times10^{-16}$ &
$\mbox{hyper, }\la_7=3\times10^{-20}$ & 3 & 13 & 28\nl
B   &  $-$ & $1\times10^{-3}$ & $\mbox{hyper, }\nu_7=5\times10^{-16}$ &
$\mbox{hyper, }\la_7=3\times10^{-20}$ & 3 & 13 & 28\nl
C   &  $-$ & $1\times10^{-5}$ & $\mbox{hyper, }\nu_7=5\times10^{-16}$ &
$\mbox{hyper, }\la_7=3\times10^{-20}$ & 3 & 13 & 28\nl
D   &  $1\times10^{-3}$ & $1\times10^{-3}$ & $\mbox{hyper, }\nu_7=5\times10^{-16}$ &
$\mbox{hyper, }\la_7=3\times10^{-20}$ &
3 & 13 & 28\nl
E   &  $5\times10^{-2}$ & $-$ & $\mbox{normal, }\nu=0.01$ & $\mbox{normal,
}\la=0.01$ & 1 & 16 & 16\nl
F   &  $5\times10^{-5}$ & $-$ & $\mbox{normal, }\nu=0.013$ & $\mbox{normal,
}\la=0.013$ & 1 & 13 & 13\nl
\enddata
\end{deluxetable}

\clearpage

\begin{figure}
\plotone{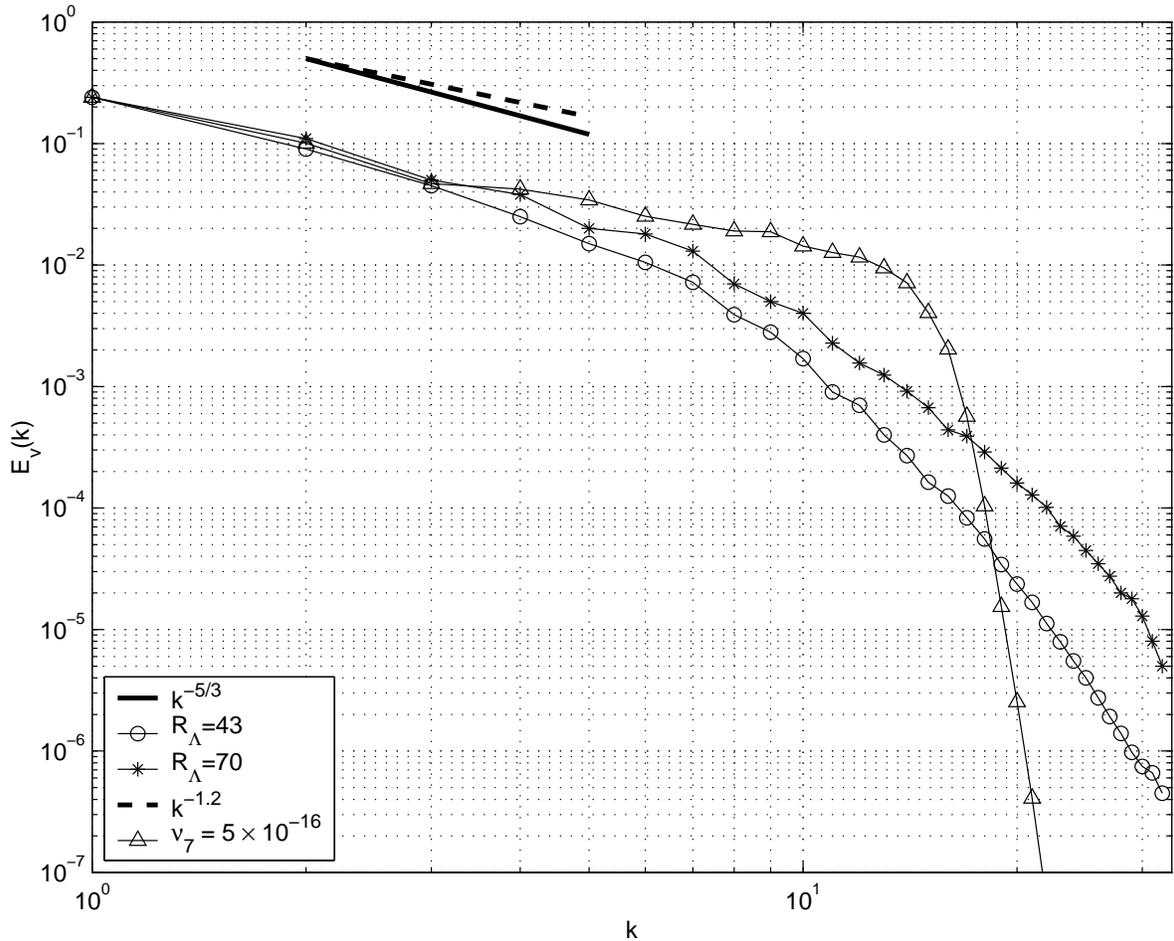}
\caption{Kinetic energy spectrum at $R_{\Lambda} = 43$ (circles) and
$R_{\Lambda} = 70$ (stars) from pure hydrodynamic turbulence simulation with a resolution of $64^3$. We used normal dissipation, $\nu \nabla^2 \U$,
for these two runs. Also shown in this figure is the kinetic energy
spectrum with a hyper-viscosity $\nu_7 = 5.0\times 10^{-16}$
(triangles), from a $64^3$ spectral simulation of pure hydrodynamic
turbulence.}
\end{figure}

\clearpage

\begin{figure}
\plotone{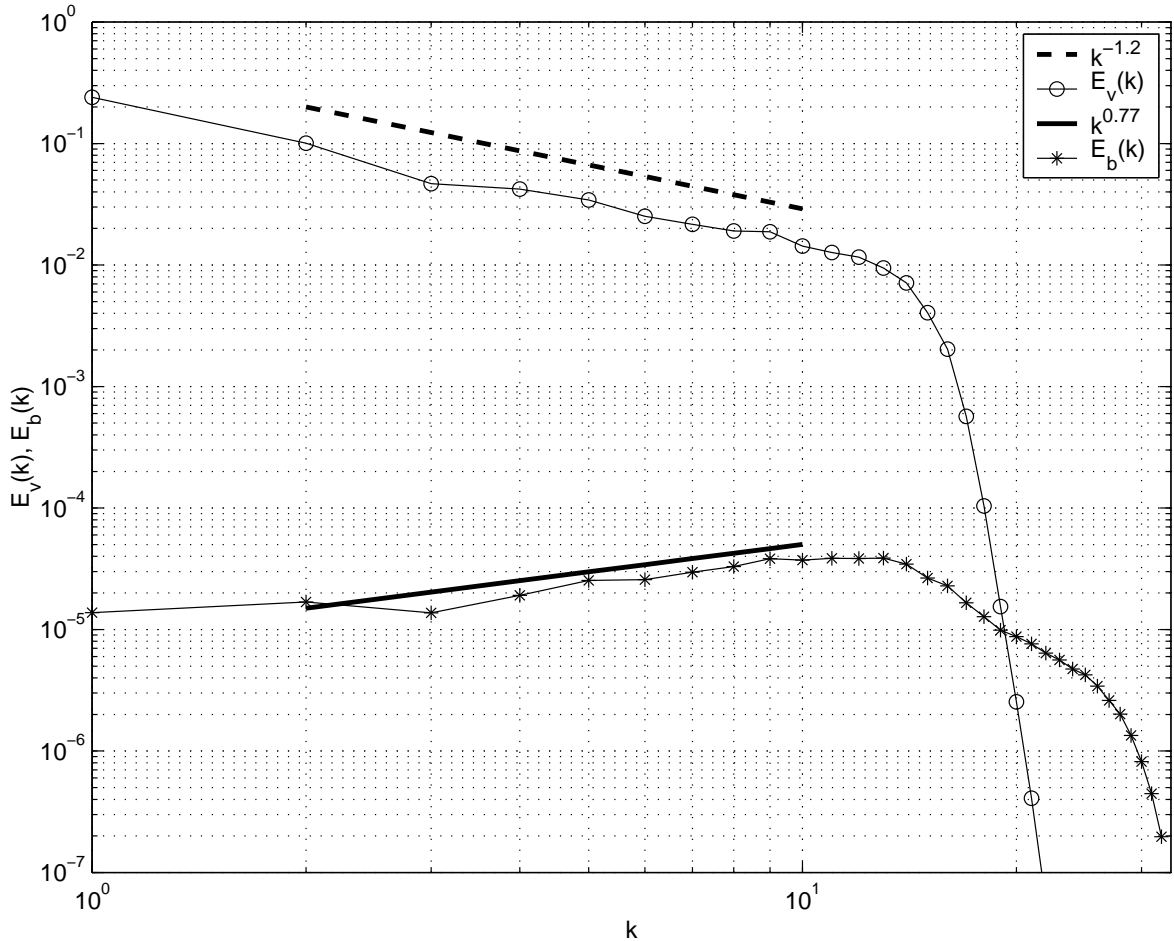}
\caption{Kinetic energy spectrum $E_v(k)$ and magnetic energy spectrum
$E_b(k)$ at $t=0.3$ for Run A, in which the initial large-scale
seed magnetic field is $\OB = {\overline B} {\hat {\bf e}}_y$ with a
${\overline B} = 0.0316$. Both axes are plotted in logarithmic
scales. For a kinetic energy spectrum of the form $k^{-p}$, the
magnetic energy spectrum, which is generated from the interaction
between the turbulence and $\OB$, follows $k^{-p+2} \sim k^{0.77}$.}
\end{figure}

\begin{figure} 
\plotone{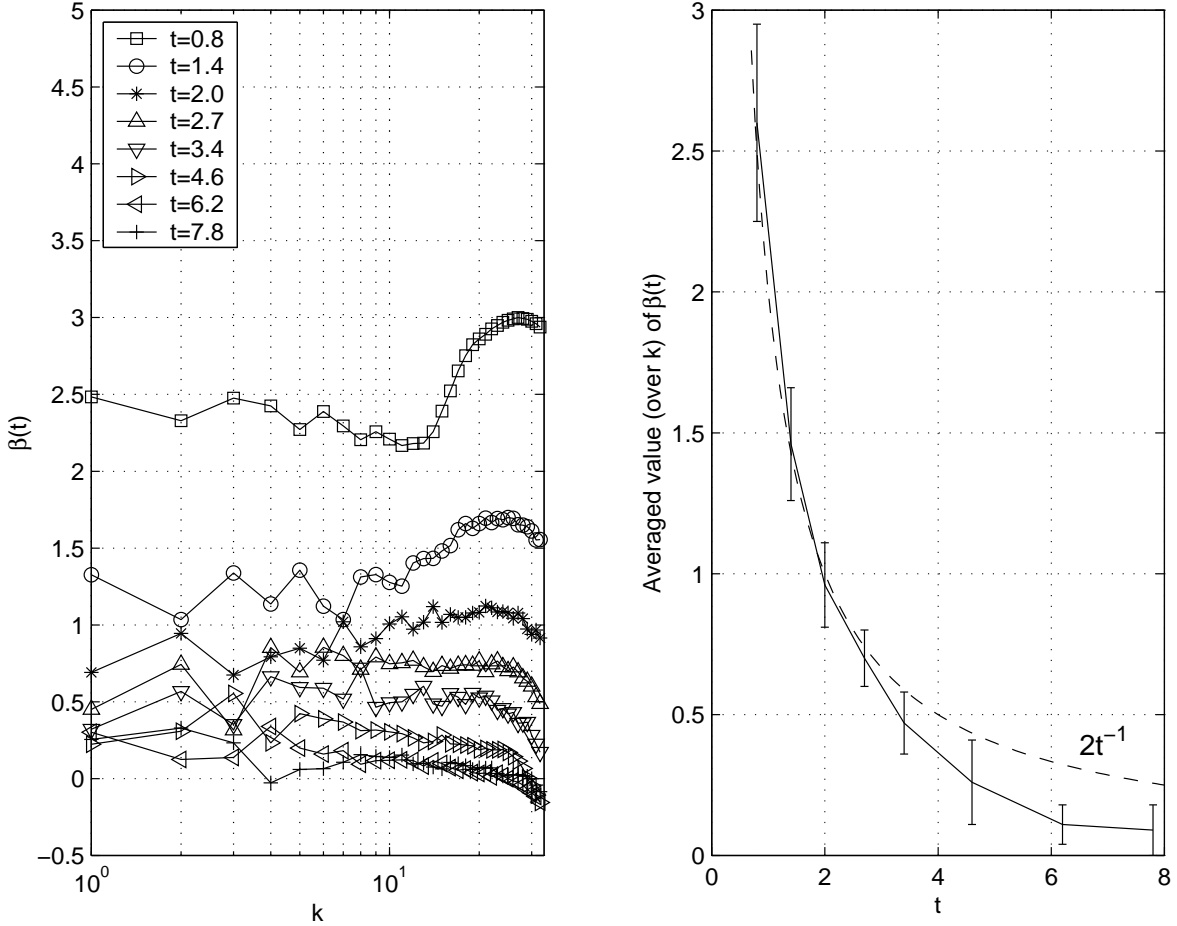}
\caption{$\beta(t)$, as defined in equation (13) in main text, at
various times are plotted in the left panel of this figure. The averaged
value (over wavenumber $k$) of $\beta(t)$ as a function of time is
plotted in the right panel. In the left panel, the horizontal axis is
plotted in logarithmic scale, while the vertical axis is plotted in
linear scale. In the right panel, both axes are plotted in linear
scales. Data shown here are from Run A, in which the initial large-scale
seed magnetic field is $\OB = {\overline B} {\hat {\bf e}}_y$ with a
${\overline B} = 0.0316$.}
\end{figure}

\clearpage

\begin{figure}
\plotone{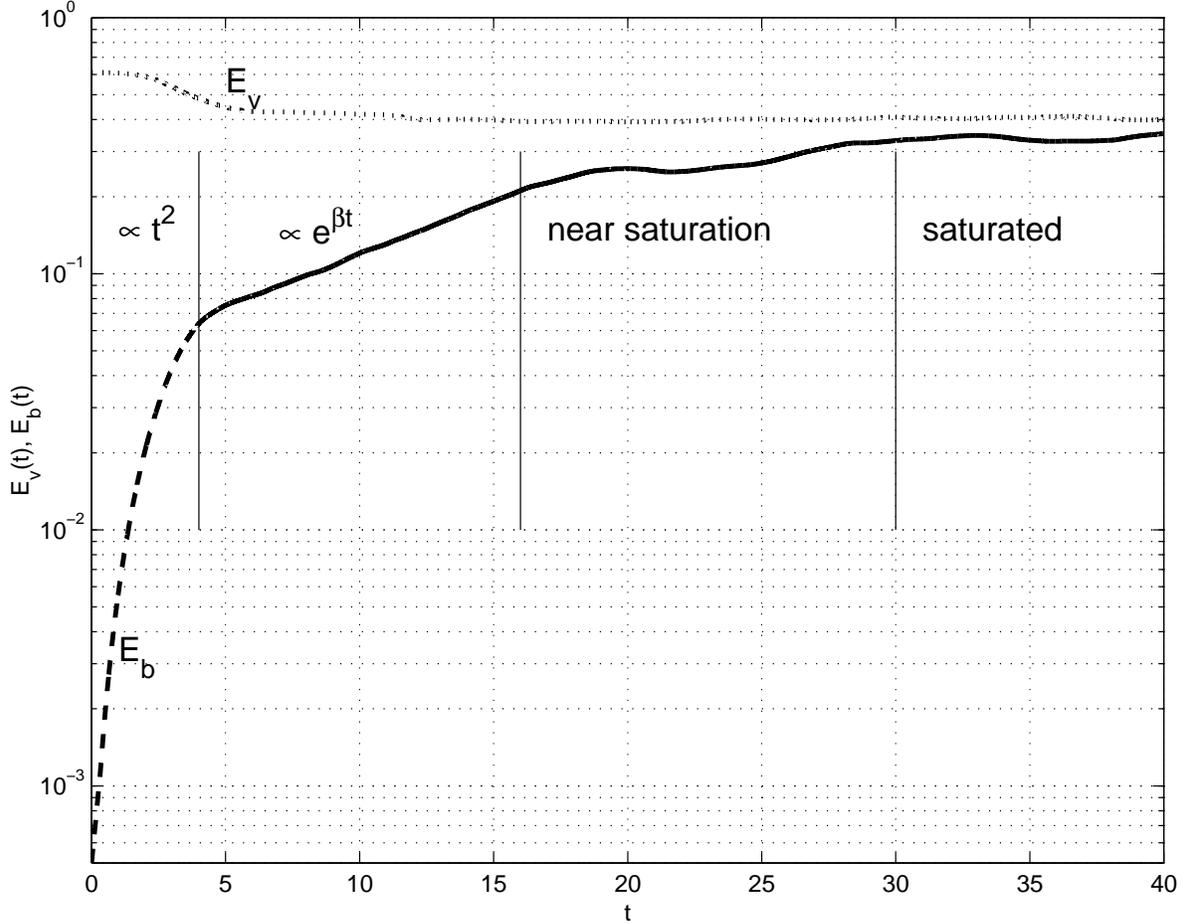}
\caption{Kinetic energy density $E_v = \sum_{k}E_v(k)$ and magnetic
energy density $E_b = \sum_{k}E_b(k)$ as functions of
time. Same conditions as Figure 3. The growth of magnetic energy
density follows four stages. Stage 1 starts from the beginning of the
simulation to $t=3$, during which the magnetic energy grows as
$t^{2}$. During stage 2 that lasts from $t=4$ to $t=16$, it grows
exponentially with growth rate $\beta = 0.1$. After the exponential
growth stage, magnetic energy density growth enters a near saturation
stage from $t=16$ to $t=30$, where the growth is further slowed
down. For the saturation stage, the magnetic energy density fluctuates
around 0.33, while the kinetic energy density fluctuates around 0.4.}  
\end{figure}

\clearpage

\begin{figure}
\plotone{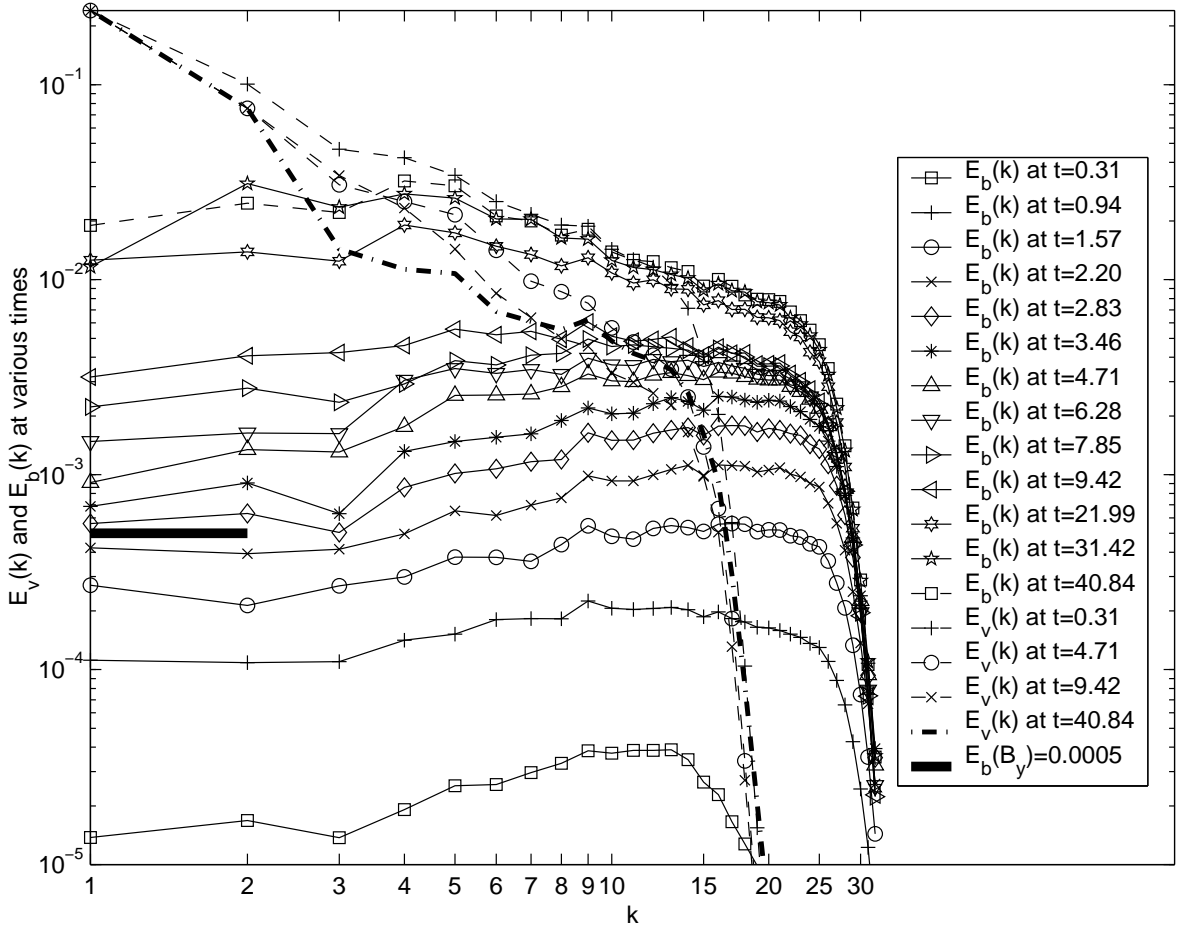}
\caption{Kinetic energy spectrum $E_b(k)$ and magnetic energy spectrum
$E_v(k)$ at different times. Data shown here are from Run A, in which
the initial large-scale seed magnetic field is $\OB = {\overline B}
{\hat {\bf e}}_y$ with a ${\overline B} = 0.0316$. The thick horizontal
line denotes the energy density due to $\OB$. Growth of the magnetic field
at all scales can be seen. The growing magnetic field suppresses
the velocity field in the first three stages defined in Figure 4.}
\end{figure}

\clearpage

\begin{figure}
\plotone{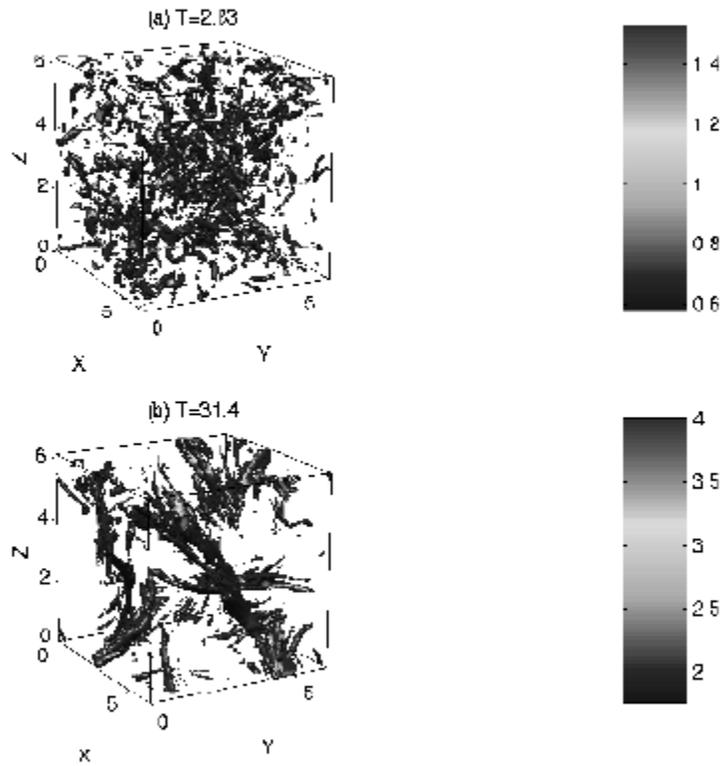}
\caption{Isosurface of magnetic field magnitude at different
times. The data are from Run A, in which the initial large-scale
seed magnetic field is $\OB = {\overline B} {\hat {\bf e}}_y$ with a
${\overline B} = 0.0316$. The isosurfaces are determined at
$2.5\times \mbox{mean}\{|B|\}$, where $\mbox{mean}\{|B|\}=0.23$ for
(a) and $\mbox{mean}\{|B|\}=0.70$ for (b). The emergence of magnetic
structures of the size of the simulation box can be seen at $t=31.40$.}
\end{figure}

\clearpage

\begin{figure}
\plotone{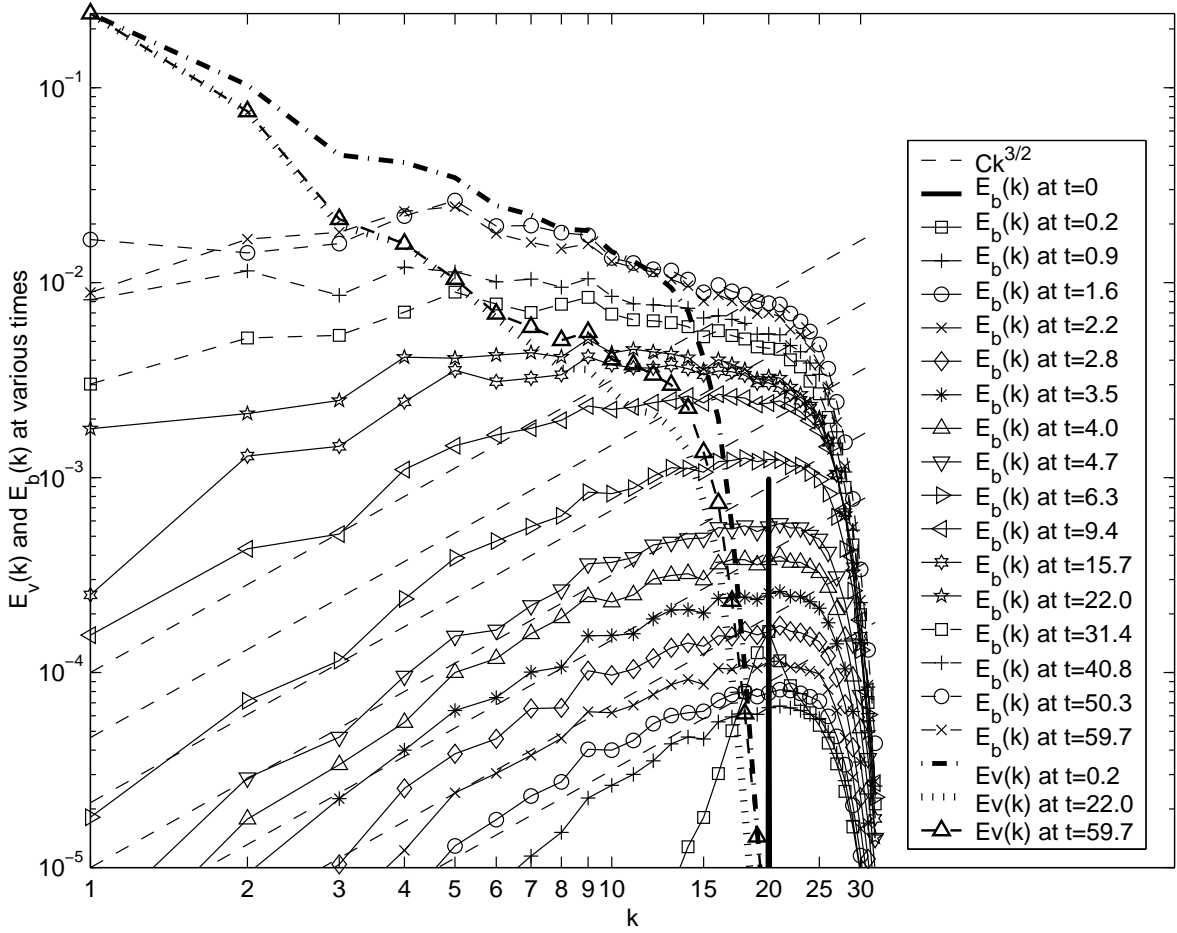}
\caption{Magnetic energy spectrum $E_b(k)$ and kinetic energy spectrum
$E_v(k)$ at various times. Data shown here are from Run B, in which an
initial seed magnetic energy is concentrated at $k=20$ with an
$e_0=1\times 10^{-3}$. Dashed lines are $Ck^{3/2}$ ($C$ varies), which is
reminiscent of the prediction by Kulsrud \& Anderson(1992) for the
magnetic spectrum that grows from a magnetic impulse. }   
\end{figure}

\clearpage

\begin{figure}
\plotone{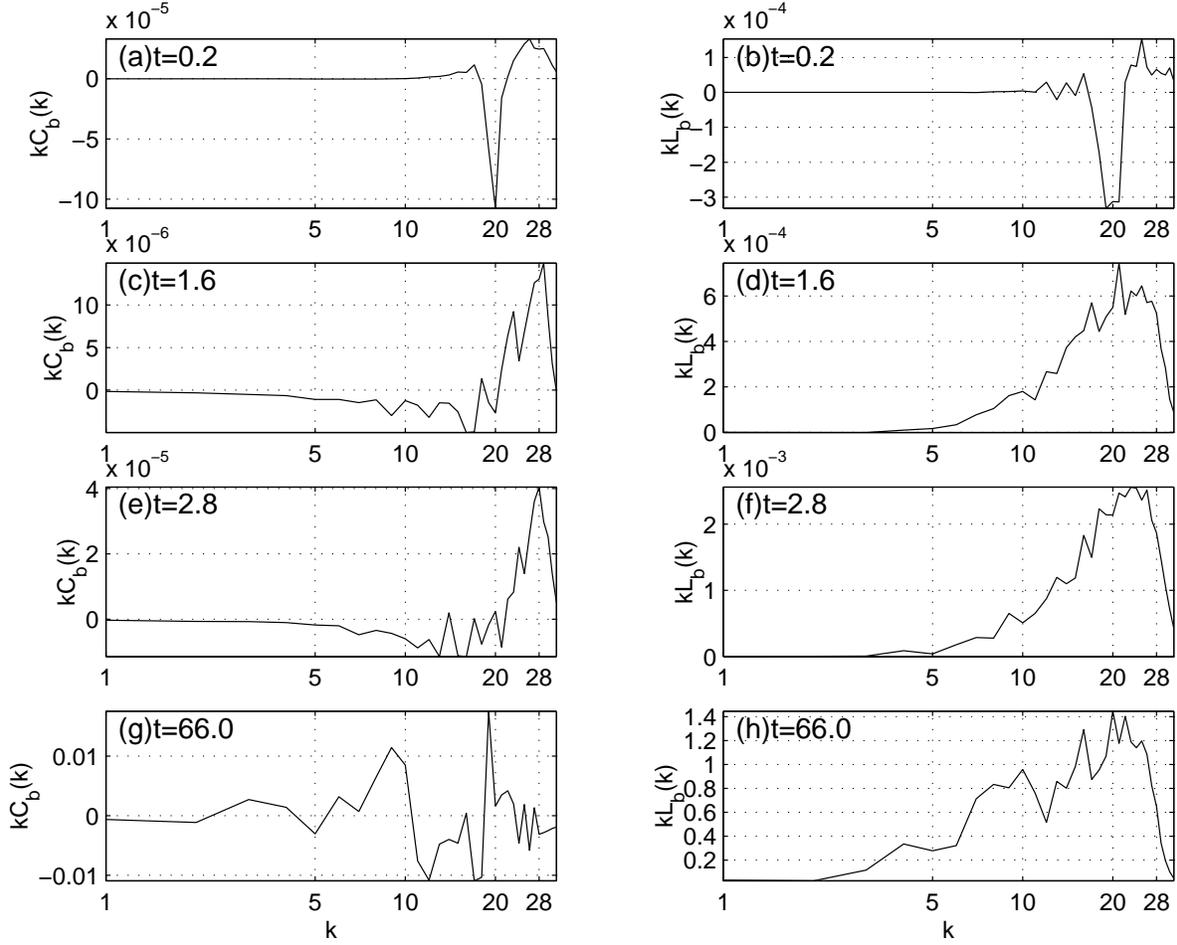}
\caption{Plots of the convective effect $kC_b(k)$ and the linear
stretching effect $kL_b(k)$ at different time
points. The horizontal axis is plotted in logarithmic scale, and the
vertical axis is plotted in linear scale. Data shown here are from Run
B, in which an initial seed magnetic energy is concentrated at $k=20$
with an $e_0=1\times 10^{-3}$. Note the different scales of $y$-axes of
different panels. }
\end{figure}

\clearpage

\begin{figure} 
\plotone{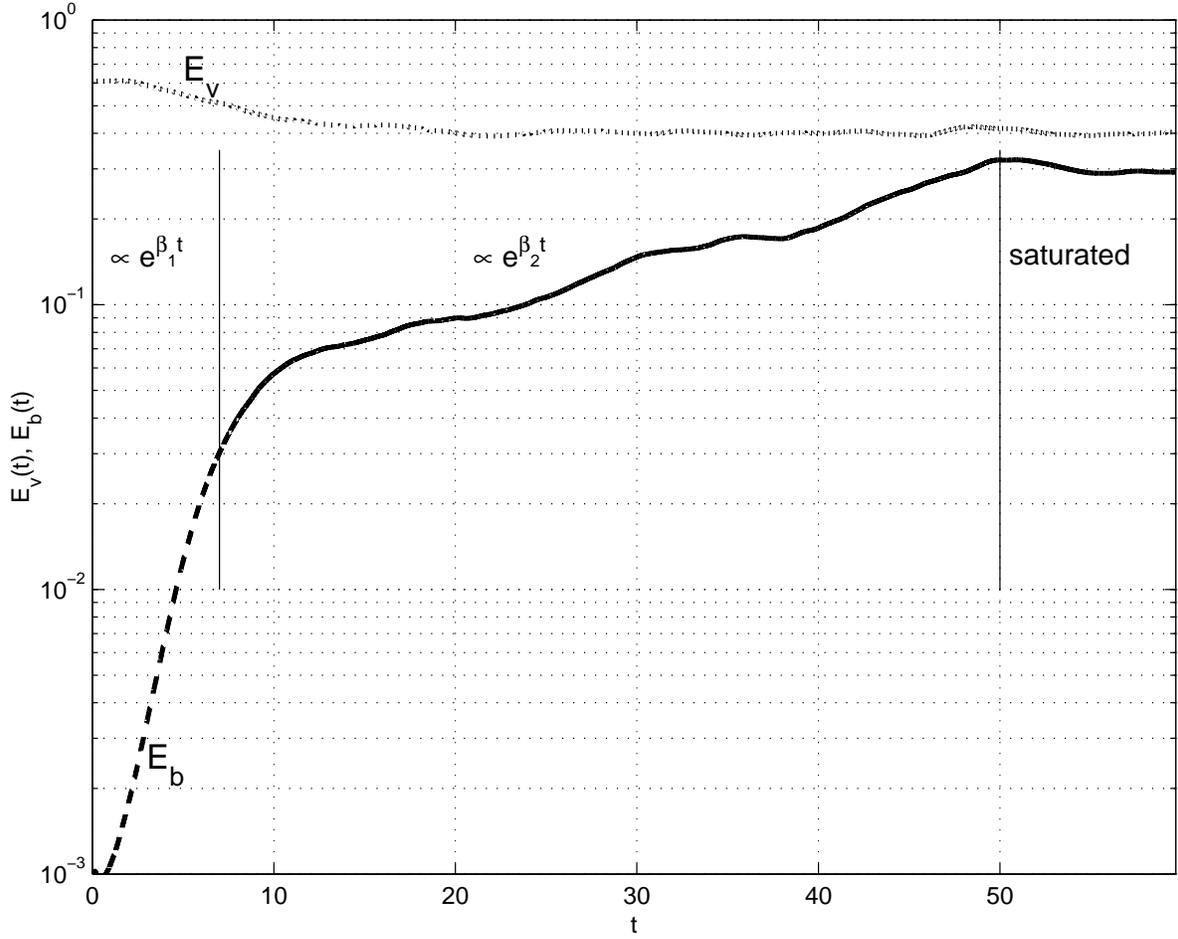}
\caption{Temporal evolution of kinetic energy density and magnetic
energy density. Data shown here are from Run B, in which an initial seed
magnetic energy is concentrated at $k=20$ with an $e_0=1\times 10^{-3}$.}   
\end{figure}

\clearpage

\begin{figure}
\plotone{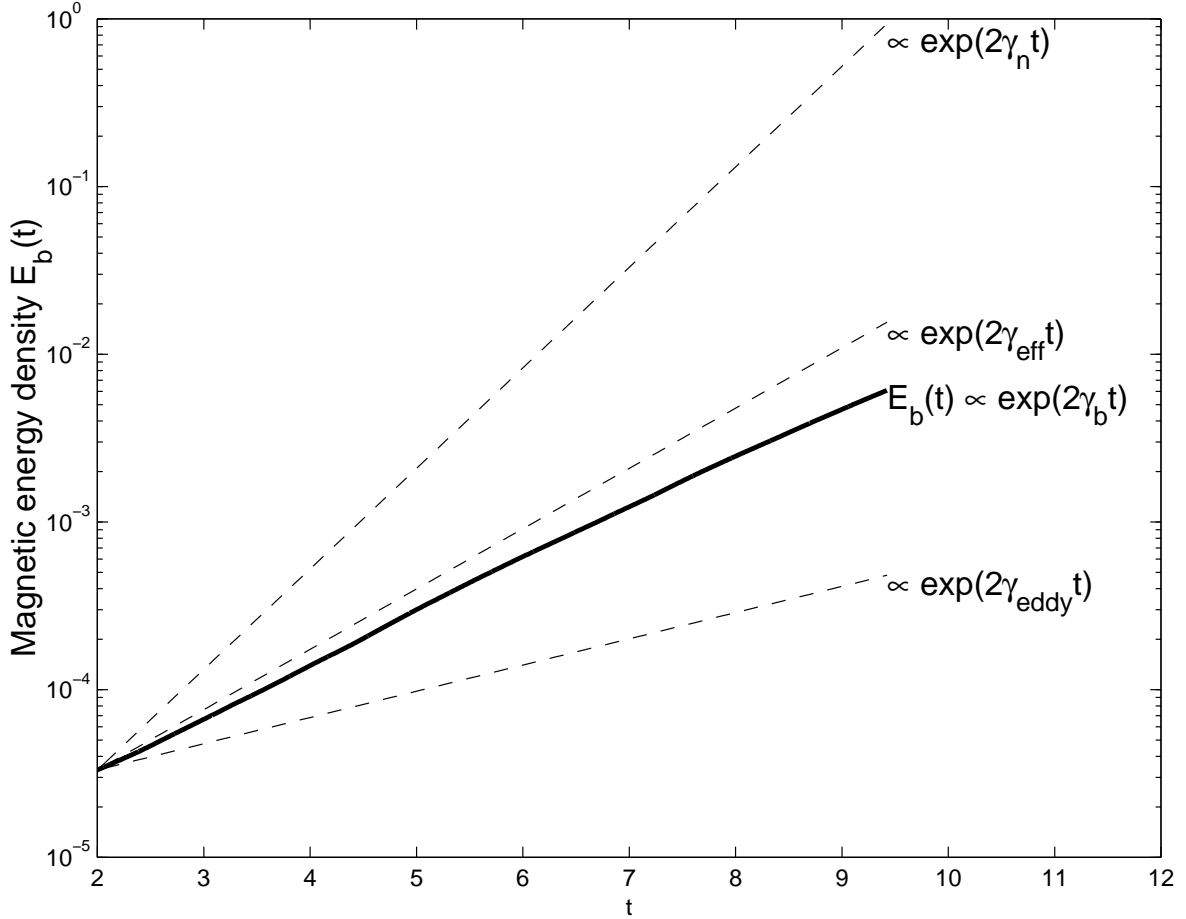}
\caption{Magnetic energy density $E_b = \sum_{k}E_b(k)$ as a function
of time during the exponential growth stage of Run C, in which an initial seed
magnetic energy is concentrated at $k=20$ with an $e_0=1\times
10^{-5}$. $\gamma_b = 0.36$. $\gamma_n = 0.69$ is the inverse of
turnover time of the smallest eddies. $\gamma_{eff} = 0.6 \gamma_n$ as
predicted by Chandran(1997) and Schekochihin \&
Kulsrud(2000). $\gamma_{eddy} = 0.18$ is the inverse of turnover time
of the largest eddies.}
\end{figure}

\clearpage

\begin{figure}
\plotone{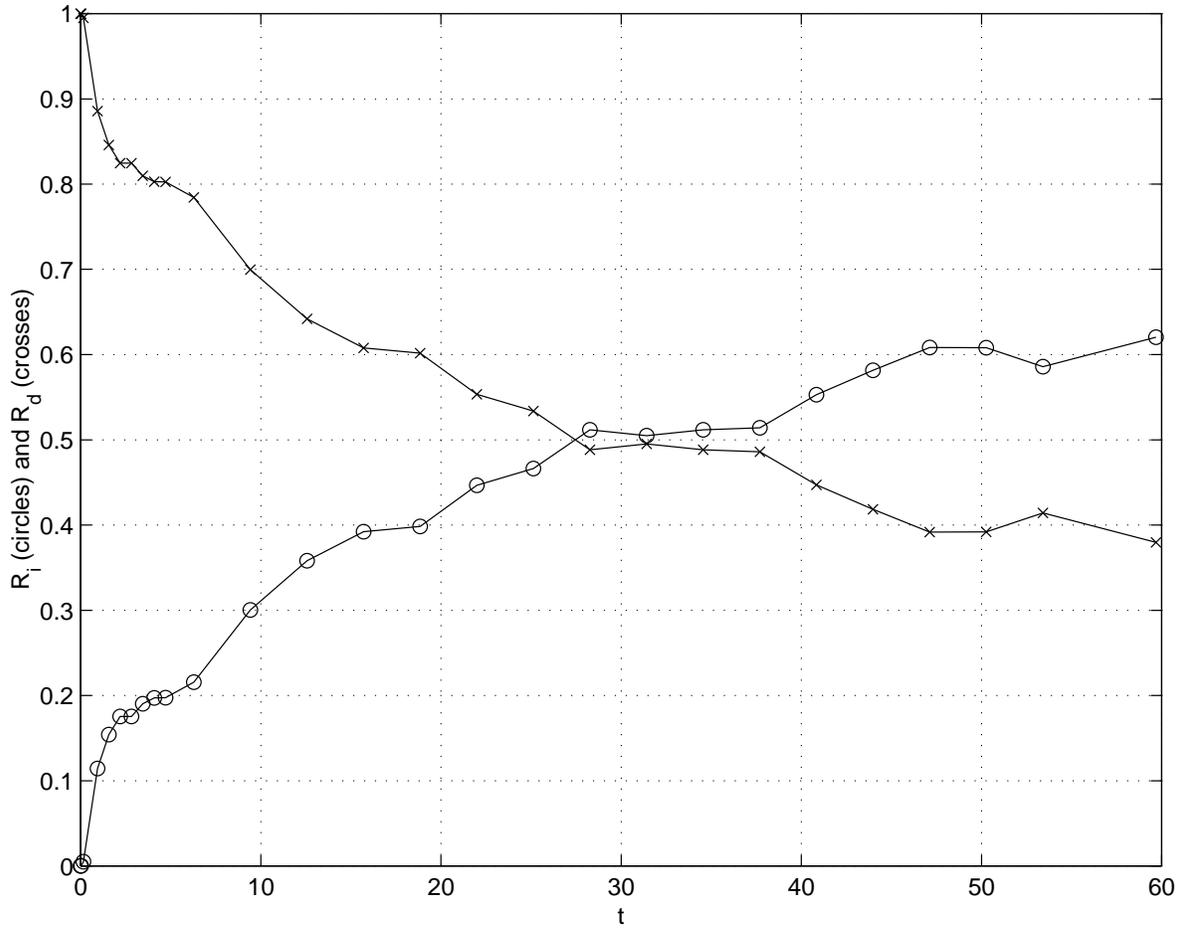}
\caption{Comparison of the magnetic energy within the inertial range and
the magnetic energy within the dissipation range. Note the
linear-linear plot scales. Data shown here are from Run B, in which an
initial seed magnetic energy is concentrated at $k=20$ with an
$e_0=1\times 10^{-3}$.}
\end{figure}

\clearpage

\begin{figure}
\plotone{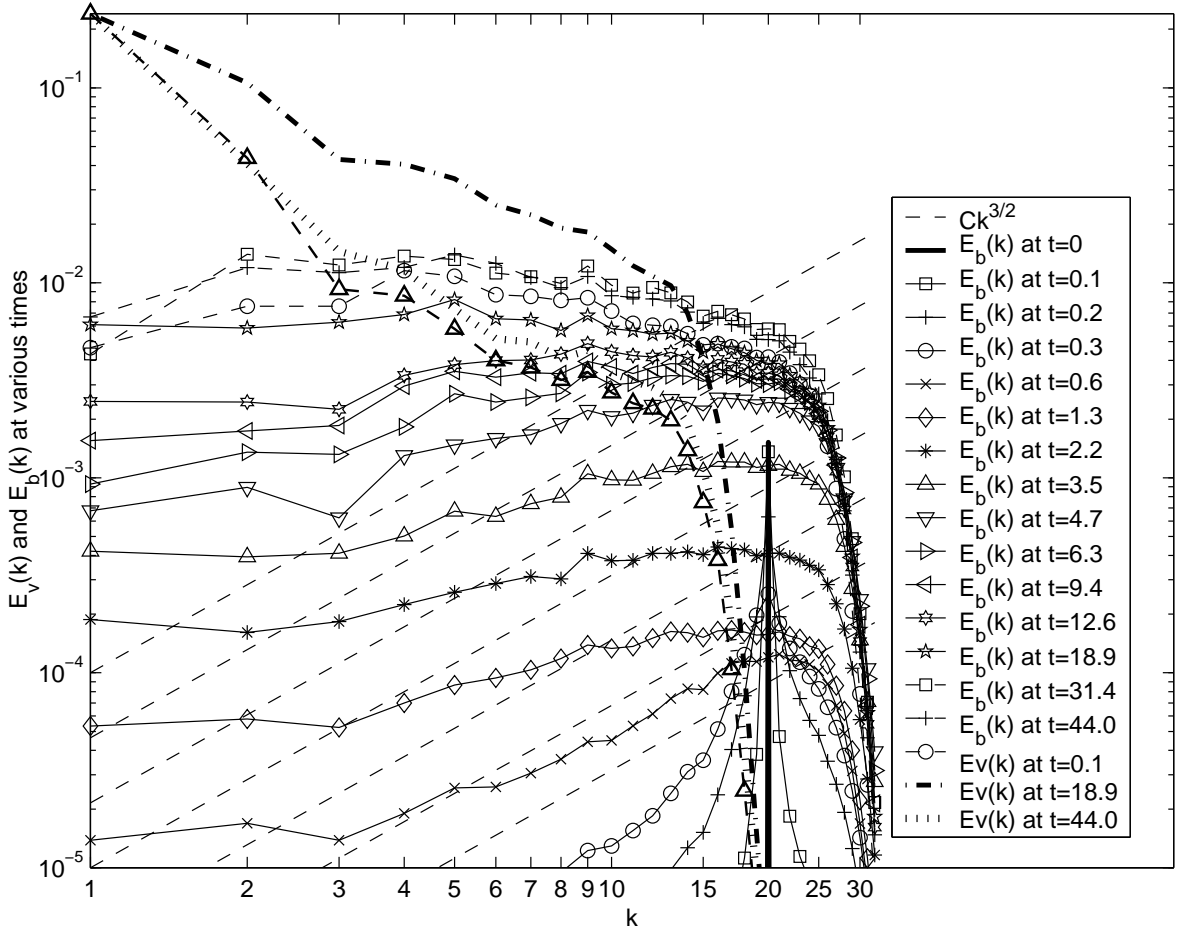}
\caption{Magnetic energy spectrum $E_b(k)$ and kinetic energy spectrum
$E_v(k)$ at different time points. Data shown here are from Run
D, in which an initial seed magnetic field is composed of a
large-scale seed field, $\OB = {\overline B} {\hat {\bf e}}_y$ with a
${\overline B}^2/2 = 1\times10^{-3}$, and a small-scale seed
field that is concentrated at $k=20$ with an $e_0=1\times
10^{-3}$. Straight dashed lines are $Ck^{3/2}$ for various values of
$C$.}
\end{figure}

\end{document}